\documentclass[pra,twocolumn,showpacs,floatfix]{revtex4}
\usepackage{graphicx}
\usepackage[usenames]{color} 
\newcommand{\beq}{\begin{equation}}
\newcommand{\eeq}{\end{equation}} 
\newcommand{\beqa}{\begin{eqnarray}}
\newcommand{\eeqa}{\end{eqnarray}} 
\newcommand{\ba}{\begin{array}}
\newcommand{\ea}{\end{array}}

\begin{document}

\title{Localization of a Bose-Einstein condensate 
in a bichromatic optical lattice}
\author{S. K. Adhikari$^{1}$\footnote{adhikari@ift.unesp.br;
URL: www.ift.unesp.br/users/adhikari}}
\author{L. Salasnich$^{2}$\footnote{luca.salasnich@pd.infn.it; 
URL: www.padova.infm.it/salasnich}}
\affiliation{$^1$Instituto de F\'{\i}sica Te\'orica, UNESP - S\~ao Paulo 
State University, Barra Funda,  01.140-070 S\~ao Paulo, S\~ao Paulo, Brazil\\
$^2$CNR-INFM and CNISM, Unit\`a di Padova,
Dipartimento di Fisica ``Galileo Galilei'', Universit\`a di Padova, Via
Marzolo 8, 35131 Padova, Italy}

\begin{abstract} By direct numerical simulation of the time-dependent 
Gross-Pitaevskii equation we study different aspects of the localization 
of a non-interacting ideal 
Bose-Einstein condensate (BEC) in a one-dimensional bichromatic 
quasi-periodic optical-lattice potential. Such a  quasi-periodic potential,  
used in a recent experiment on the localization of a BEC [Roati {\it et 
al.}, Nature {\bf 453}, 895 (2008)], can be formed by the 
superposition of two standing-wave polarized 
laser beams with different wavelengths. We investigate the effect of 
the variation of optical amplitudes and wavelengths on the localization 
of a non-interacting BEC. We also simulate the non-linear 
dynamics when a harmonically trapped BEC is suddenly released 
into a quasi-periodic potential, { as done experimentally 
in a laser speckle potential [Billy {\it et al.}, Nature {\bf 453}, 891 (2008)]}.
We finally study the destruction of the localization in an interacting BEC
due to  the repulsion generated by a positive scattering length between 
the bosonic atoms. 
\end{abstract}


\maketitle

\section{Introduction}

The localization of the electronic wave 
function in a disordered potential was predicted by 
Anderson fifty years ago \cite{anderson}. 
More recently the phenomenon of localization due to 
disorder was experimentally observed in electromagnetic 
waves \cite{light,micro}, in sound waves \cite{sound}, 
and also in quantum matter waves \cite{billy,roati,chabe,edwards}. 
In the case of quantum matter waves, 
Billy {\it et  al.} \cite{billy} observed exponential localization of 
{ a Bose-Einstein condensate} (BEC) of $^{87}$Rb atoms 
released into a one-dimensional (1D) waveguide in the 
presence of a controlled disorder created by a laser speckle. 
Roati {\it et  al.} \cite{roati} observed localization of a 
non-interacting BEC of $^{39}$K atoms in a 1D potential created by 
two optical-lattice (OL) potentials with different amplitudes 
and wavelengths. The non-interacting BEC of 
$^{39}$K atoms was created \cite{roati} by tuning the inter-atomic 
scattering length to zero near a Feshbach resonance \cite{fesh}. 

In this paper, with intensive numerical simulations 
of the {\it linear}
Gross-Pitaevskii (GP) equation, (which is just the Schr\"odinger equation,) we study 
different aspects of localization 
{of the BEC}
in a 1D 
bichromatic quasi-periodic OL potential used in the 
experiment of Roati {\it et al.} \cite{roati}. The 
1D quasi-periodic potentials have a spatial ordering that is 
intermediate between periodicity and disorder \cite{harper,aubry,thouless}. 
In particular, the 1D discrete Aubry-Andre model of quasi-periodic 
confinement \cite{aubry,thouless} displays a transition from extended 
to localized states which resembles the Anderson localization 
of random systems \cite{random,optical}. Modugno \cite{modugno} 
has recently shown that the linear 1D Schr\"odinger equation 
with a bichromatic periodic potential 
can be mapped in the Aubry-Andre model 
and {he}  studied the transition to localization 
as a function of the parameters of the periodic potential. 
 
To investigate the interplay between the bichromatic potential 
and the inter-atomic interaction in the localization of a BEC, 
we adopt the 1D {\it non-linear} GP equation \cite{book,GP} in place of 
the linear 1D Schr\"odinger equation used to describe a non-interacting BEC. 
We find that the non-linearity of the GP equation, which accurately 
models the binary inter-atomic interaction of atoms,  
has a strong effect on localization and a reasonably 
weak repulsive non-linear 
term is capable of destroying the localization. 
Our results on the  effect of  non-linearity in the 
localization are thus in qualitative agreement with similar predictions 
based on the 1D discrete non-linear Schrodinger equation (DNLSE)
with random on-site energies \cite{dnlse}. 
First effects of a weak non-linearity in Anderson
localization have been shown experimentally in light waves in photonic
crystals \cite{roati,lahini}.

There have already been a number of { theoretical} 
studies on Anderson localization 
 under the action of different potentials. Sanchez-Palencia {\it et
al.} and Cl\'ement {\it et
al.} considered Anderson localization in a random potential
\cite{random}. Damski {\it et al.} and Schulte {\it et al.}
 considered Anderson localization in disordered OL
potential \cite{optical}. There have been studies of Anderson
localization with other types of disorder \cite{other}.
Effect of interaction on Anderson localization was also studied
\cite{interaction}.
Anderson localization in BEC under the action of a disordered potential
in two and three dimensions has also been investigated \cite{2D3D}.
{In this paper we study different aspects of the localization of an ideal BEC
and the delocalization of an interacting BEC in a quasi-periodic OL potential 
using the linear and non-linear 1D GP equation. }

In Sec. \ref{II} we present a brief account of the non-linear 1D GP 
equation used in our study and of the variational solution of the same 
under appropriate conditions. In Sec. \ref{III} we present our numerical 
studies on localization using time propagation under the Crank-Nicolson 
discretization scheme. The density profile of the tightly localized 
states are in agreement with the variational results. We also study the 
effect of the variation of the wavelength and intensity of the OL 
potentials on the localization. We study the non-equilibrium  dynamics, 
as observed in the experiment of Billy {\it et al.} \cite{billy}, when a 
harmonically trapped BEC is suddenly released from the harmonic trap 
into a quasi-periodic OL potential. We also investigate the destruction 
of localization in the presence of a repulsive atomic interaction and it 
is found that a reasonably small non-linearity can destroy the 
localization of the BEC. In Sec. \ref{IIII} we present a brief 
discussion and concluding remarks.

\section{Analytical consideration of localization}

\label{II}

In the actual experiment of Roati {\it et al.} \cite{roati}
the 1D quasi-periodic  bichromatic OL potential was 
produced by superposing two OL potentials generated by two 
standing-wave polarized laser beams of slightly different wavelengths 
and amplitudes, which we take here similar to 
those in the experiment of Roati {\it et al.}, e.g., with wavelengths 
$\hat \lambda_1=1032$ nm and $\hat \lambda_2=862$ nm. This 
1D quasi-periodic OL  potential can be written  as  \cite{roati} 
\begin{equation}
\label{pot}
V( \hat z)=\sum_{i=1}^2 {\color{Red}2}s_iE_{i}\cos^2(k_i \hat z),
\end{equation}
where ${\color{Red}2}s_i, i=1,2,$ are the amplitudes  of the OL potentials in units of 
respective recoil energies $E_i=2\pi^2 \hbar^2/(m \hat \lambda_i^2)$, and 
$k_i=2\pi/\lambda_i$, $i=1,2$ are the respective wave numbers, $\hbar$ 
is the reduced Planck constant, and $m$ the mass of an atom. 

With  a single periodic potential of the form $\cos^2(k \hat z)$ 
with $s_2=0,$  
the linear Schr\"odinger equation permits only 
de-localized states in the form of Bloch waves. Localization is possible 
in the linear Schr\"odinger equation due to the ``disorder'' introduced 
through a second component in Eq. (\ref{pot}). The localization is not 
intuitively obvious. The potential (\ref{pot}) continue 
to have an infinite number of finite barriers as in a simple OL 
potential, and it might be expected that any localized state will decay 
due to tunneling. 

The localized states that we study are { low-lying states} 
of the system with potential (\ref{pot}). They are quite 
distinct from the so called gap solitons in a simple OL potential
appearing for repulsive non-linearity 
in the band-gap of the spectrum of the linear Schr\"odinger 
equation \cite{gs}. These gap solitons with finite spatial extension 
are {excited  states} of the system { without a linear 
counterpart}. 

The  dynamics of a trapped BEC of $N$ atoms 
in a transverse harmonic potential of frequency $\omega_{\bot}$
plus the axial quasi-periodic OL potential (\ref{pot}) 
is determined by the following GP equation \cite{book,GP}
\begin{eqnarray}
\label{e2}
i\frac{\partial }{\partial  t} \psi({\bf \bf r},t) &=& 
\biggr[ -\frac{\nabla^2}{2}+\frac{x^2+y^2}{2} 
+V(z)\nonumber \\ & +& 2\pi  g |\psi({\bf r},t)|^2 \biggr]\psi({\bf \bf 
r},t) \; , \\
V(z)&=& \sum_{i=1}^{2}\frac{4\pi^2s_i}{\lambda_i^2}\cos^2\biggr(
\frac{2\pi}{\lambda_i}
z\biggr),\label{pot1}
\end{eqnarray}
with normalization $\int |\psi({\bf \bf
r},t)  |^2 d{\bf r}=1$ and 
where 
$g=2N\hat a/a_{\bot}$ is the dimensionless interaction strength 
with $\hat a$ the inter-atomic scattering length and 
$a_{\bot}=\sqrt{\hbar/(m\omega_{\bot})}$ the characteristic 
harmonic length of the transverse confinement, and ${\bf r}\equiv
(x,y,z)$ defines the Cartesian coordinates. 
Here we have considered the harmonic trap $(x^2+y^2)/2$ 
in transverse directions $(x,y)$ and the quasi-periodic potential $V(z)$
in the longitudinal direction $z$.
 In Eq. (\ref{e2}) length is in units of $a_{\bot}$, 
time in units of $\omega_{\bot}$, and energy in units 
of $\hbar\omega_{\bot}$. In the non-interacting case $g=0$ and Eq. (\ref{e2})
becomes the usual linear Schr\"odinger equation.

Another completely equivalent potential is the one where the
cosine term of Eq. (\ref{pot1}) is replaced by a sine: 
\begin{equation}
V(z)=\sum_{i=1}^{2}\frac{4\pi^2s_i}{\lambda_i^2}\sin^2\biggr(
\frac{2\pi}{\lambda_i}
z\biggr).\label{pot2}
\end{equation}
However,  potential (\ref{pot2}) generates a different type of
localized states
compared to potential  (\ref{pot1}). Potential  (\ref{pot2}) has a local
minimum at the center $z=0$, consequently stable stationary solutions
with this potential have a maximum at $z=0$. However, potential
(\ref{pot1}) has a local maximum at $z=0$ corresponding to a minimum of
the stationary solution at the center. 
{ We shall show that, starting with an initial Gaussian wave function 
centered at $z=0$, the numerical solution of the GP equation (\ref{e2}) 
gives different localized eigenstates depending on the choice of 
the confining potential}.  

For a cigar-shaped trap with strong transverse confinement, it is 
appropriate to consider a 1D reduction of Eq. (\ref{e2}) by freezing the 
transverse dynamics to the respective ground state and integrating over 
the transverse variables. The resulting 3D-1D reduction of Eq. (\ref{e2}) 
for small non-linearity $g$ is realized via \cite{S1,M1} 
\begin{eqnarray}
\label{gp}
i\frac{\partial }{\partial t} \phi(z,t) = 
\biggr[-\frac{\partial_z^2}{2}+ V(z)
+g|\phi(z,t)|^2\biggr]\phi(z,t)  \; ,  
\end{eqnarray}
with normalization $\int_{-\infty}^{\infty} dz |\phi(z,t)|^2 =1   $  and 
where $\phi(z,t)$ is the axial wave function of the Bose condensate. 
Eq. (\ref{gp}) is a 1D non-linear Schr\"odinger equation 
with cubic non-linearity. 

Modugno used potential (\ref{pot2}) in his study
and in addition took $\lambda_1=2\pi$ and defined 
$s_i'=2 s_i$, and 
$\beta=\lambda_1/\lambda_2\equiv 2\pi /\lambda_2$, so that (viz. Eq. 
(16) of Ref. 
\cite{modugno})
\begin{eqnarray}\label{potx}
V(z)= \frac{1}{2}s_1'\sin^2(z)+\frac{1}{2}s_2'\beta^2\sin^2(\beta z).
\end{eqnarray}
Eq. (\ref{gp}) with potential (\ref{potx}) can be  mapped  
\cite{modugno}  
in a non-linear version of the Aubry-Andre model \cite{harper,aubry} 
by expanding the wave function 
$\phi(z,t)=\sum_j c_j(t) W_j(z)$ over 
a set of orthonormal Wannier states $W_j(z)$, where 
$W_j(z)$ is maximally localized at the $j$-th minimum of the 
primary lattice. In this way one finds that the complex 
coefficients $c_j(t)$ satisfy the time-dependent 
discrete non-linear Schr\"odinger equation \cite{modugno}
\beqa 
i {d\over dt} c_j(t) &=& -J (c_{j+1}(t) + c_{j-1}(t)) \nonumber 
\\
&+& \Delta \cos{(2\pi \beta j)} c_j(t) + {\tilde g} |c_j(t)|^2 c_j(t) 
\eeqa
where  
$|c_j(t)|^2$ gives the probability of occupation 
of the $j$-th site at time $t$. 
The hopping term $J$ and the ``disorder'' term $\Delta$ 
are connected to the parameters of the bichromatic potential (\ref{pot}) 
and can be { calculated by using 
the Wannier functions} \cite{modugno,boers}.
%
%
Modugno \cite{modugno} showed that 
the linear ($g=0$) GP equation (\ref{gp}) displays localization 
for a large enough $\Delta/J$ and the localization increases as 
$\Delta/J$ increases for a wide range of values of $\beta$. 

The Fourier transformation $f(k)$
of the Anderson localized state $\phi(z)$ is
defined as
\begin{equation}\label{mom}
f(k)= \int_{-\infty}^{\infty} \exp(i2\pi kz) \phi(z),
\end{equation}
and the momentum distribution of the localized state $P(k) \equiv
f^2(k)$.

Usually the stationary bound states formed with quasi-periodic OL
potentials (\ref{pot1}) and (\ref{pot2}) occupy many sites of the
quasi-periodic OL potential and have many local maxima and minima. For
certain values of the parameters potential (\ref{pot2}) leads to
bound states confined practically to the central site of the
quasi-periodic OL potential. When this happens, a variational
approximation
with Gaussian ansatz leads to a reasonable prediction for the bound
state.

The stationary form of the linear 
Schr\"odinger equation (\ref{gp}) 
(with
$i\partial /\partial t$ replaced by a chemical potential 
$\mu$)
with
potential (\ref{pot2}) can
be derived from the
following Lagrangian
\begin{eqnarray}\label{lag}
L=\int_{-\infty}^\infty \left[ \mu \phi^2(z)
-\{\phi'(z)\}^2/2-V(z)\phi^2(z)\right]
dz -\mu,
\end{eqnarray}
by demanding $\delta L/\delta \phi = \delta L/\delta \mu=0$, 
where $\mu$
is the  prime denotes space derivative. To apply the variational
approximation we use the
Gaussian ansatz \cite{PG}
\begin{equation}\label{ans}
\phi(z)=\pi^{-1/4}\sqrt{\frac{\cal N}{w}}\exp\left(- \frac{z^2}{2w^2}
\right),
\end{equation}
where variational parameters are the norm $\cal N$, width $w$, and 
$\mu$.
This ansatz implies that the center of the stationary state  is placed
at the local
minimum at $z=0$ of the quasi-periodic OL potential (\ref{pot2}). The
substitution
of ansatz (\ref{ans}) in Lagrangian (\ref{lag}) leads to
\begin{equation}\label{lag2}
L=\mu({\cal N}-1)-\frac{\cal N}{4w^2}
+\sum_{i=1}^2\frac{A_i {\cal N}
 }{2}[
\exp(-\alpha_i^2
w^2)-1],
\end{equation}
where $A_i=4\pi^2s_i/\lambda_i^2, \alpha_i=2\pi/\lambda_i.$
The first variational equation from Eq. (\ref{lag2}),
$\partial L/\partial \mu=0,$  yields ${\cal N}=1$, which will be used
in other variational equations. The second variational equation
$\partial L/\partial w=0,$  yields
\begin{equation}\label{cf}
1=\sum_{i=1}^2 2\alpha_i^2 A_i w^4 \exp(-\alpha_i ^2 w^2),
\end{equation}
  and determines the width $w$.
The last variational equation $\partial L/\partial {\cal N}=0,$ yields
\begin{equation}\label{energy}
\mu = {1}/({4w^2})-\sum_{i=1}^2{A_i}[\exp(-\alpha_i^2 w^2)-1]/2,
\end{equation}
 which defines the chemical potential. 
(However, some caution should be 
exercised in using these variational equations: they predict a {\it 
false} bound state for a one-term periodic potential with $s_2=0$, cf. Eq. 
(\ref{cf}), and the one-term 
 potential  is known to support no localized bound state.)

\section{Numerical Results}

\label{III}

To perform a systematic numerical study of localization with potentials 
(\ref{pot1}) and (\ref{pot2})
we vary $\lambda_1$ and $\lambda_2$ maintaining 
the ratio  $\lambda_2/\lambda _1 =0.86$ (roughly the same ratio 
$\lambda_2/\lambda_1$ as in the experiment of Roati et al. 
\cite{roati}). We consider a transverse harmonic oscillator length 
$a_\perp \approx 1 $ $\mu$m, so that the experimental wavelengths 
$\hat \lambda_1 =1032$ nm and $\hat \lambda_2 =862$ nm in dimensionless 
units become $\lambda_1\approx 1$ and $\lambda_2 \approx 0.86$.
In the first part of the study we  also set scattering length $\hat a=0$ 
corresponding to ideal non-interacting bosons.  

We perform {the} numerical simulation 
employing mostly real-time propagation  with 
Crank-Nicholson discretization scheme \cite{bo}
with space step 0.025, time step 0.0005.  Imaginary-time 
propagation routine can determine the strongly 
localized  state confined to a single OL site or so
in an efficient fashion. However, 
imaginary-time propagation routine demonstrated 
difficulty for weakly-confined state extending over a large number of 
OL sites. 

\begin{figure}
\begin{center}
\includegraphics[width=\linewidth]{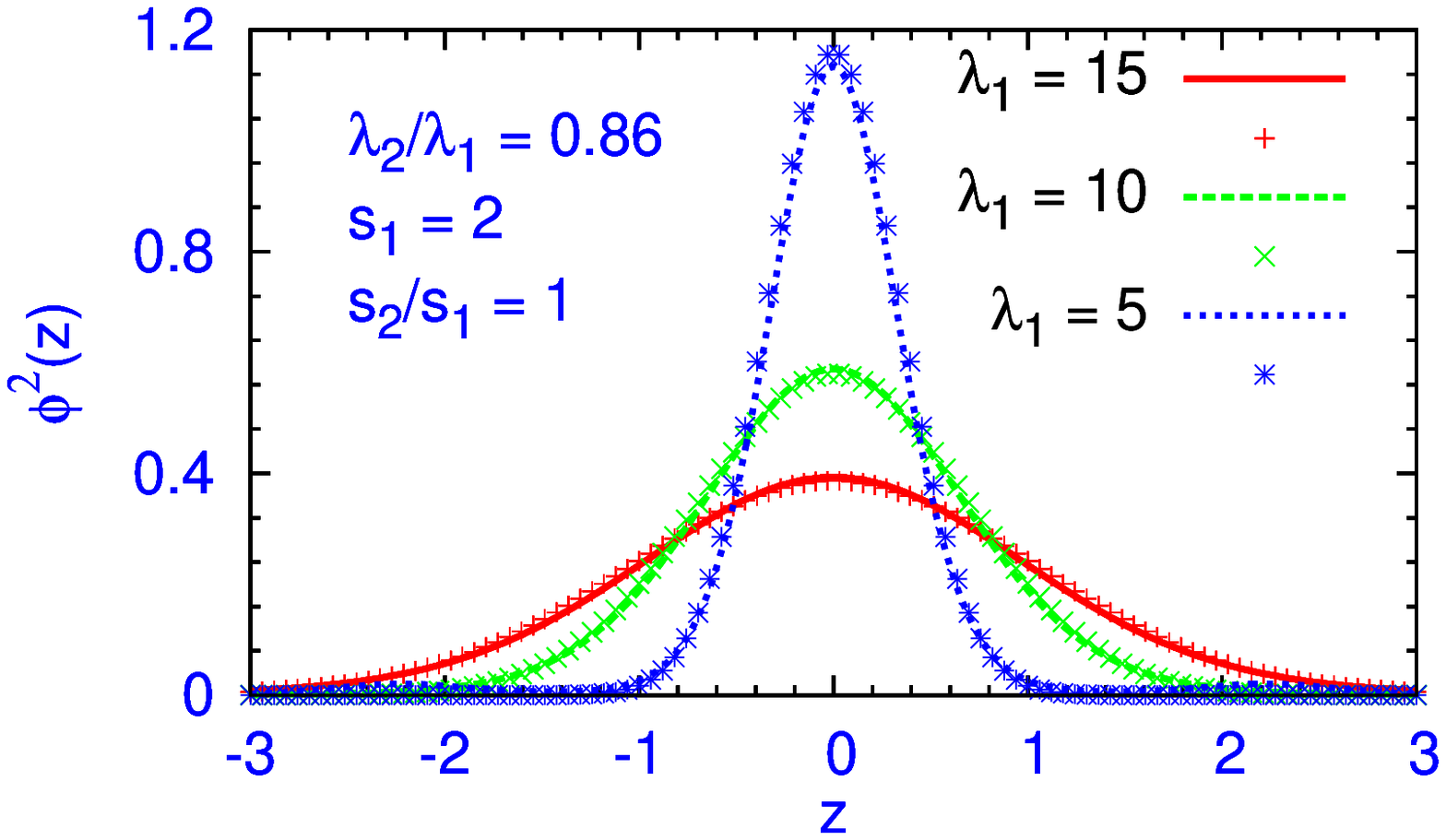}
(a)
\includegraphics[width=\linewidth]{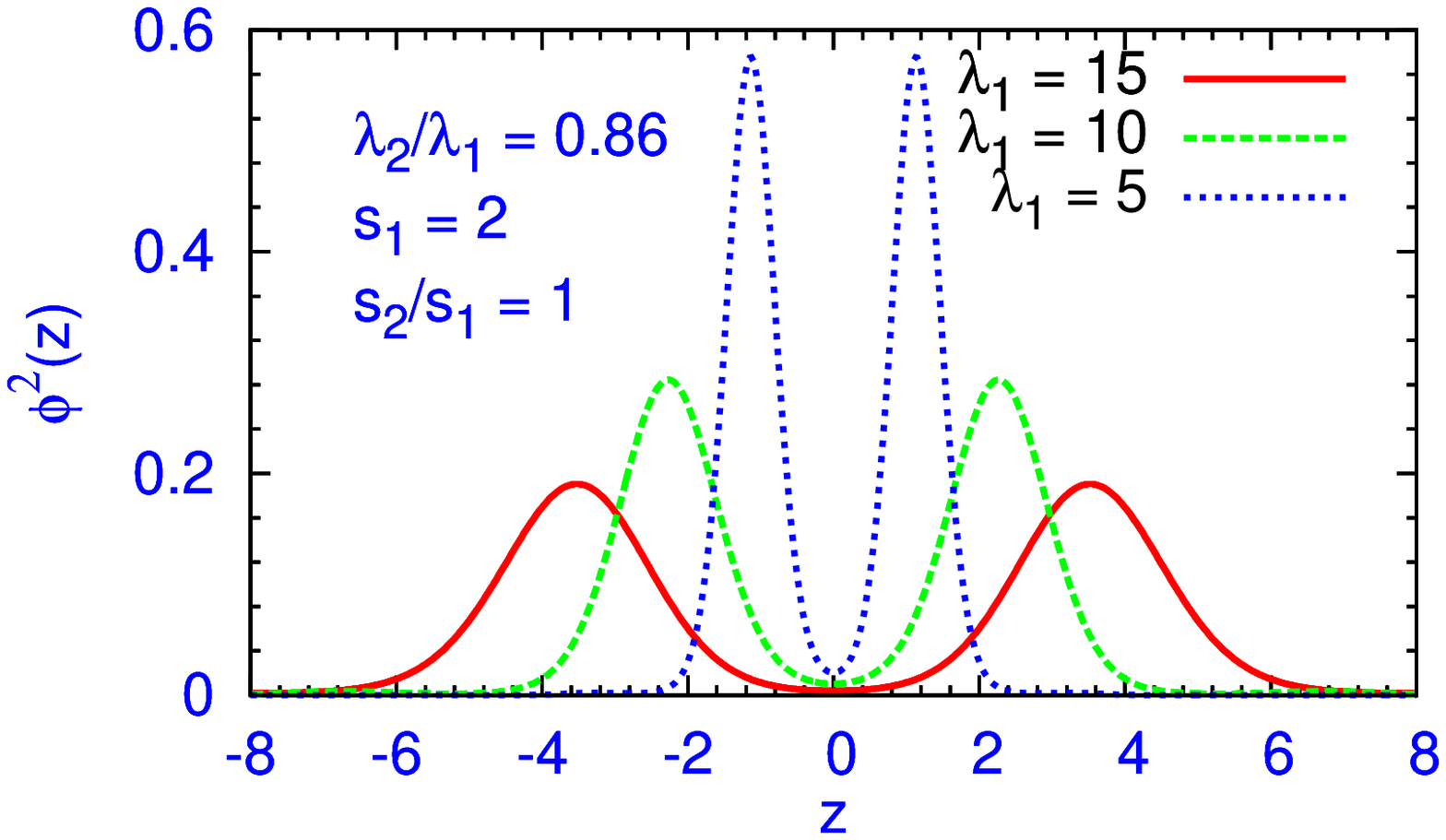}
(b)
\end{center}

\caption{(Color online) (a)  Typical density distribution $\phi^2(z)$ vs.
$z$ for   a non-interacting BEC for potential  (\ref{pot2}) 
for $\lambda_2/\lambda_1=0.86, s_1=s_2=2, $
and $\lambda_1=15, 10$ and 5. The numerical results are shown by
continuous lines. The variational results are  shown by a
chain of symbols. 
(b)  Typical density distribution $\phi^2(z)$ vs.
$z$ for potential  (\ref{pot1}) for same parameters as in (a).
Both  $\phi^2(z)$ and $z$ are   in dimensionless units.}
\label{fig1}
\end{figure}

\begin{figure}
\begin{center}
\includegraphics[width=.9\linewidth]{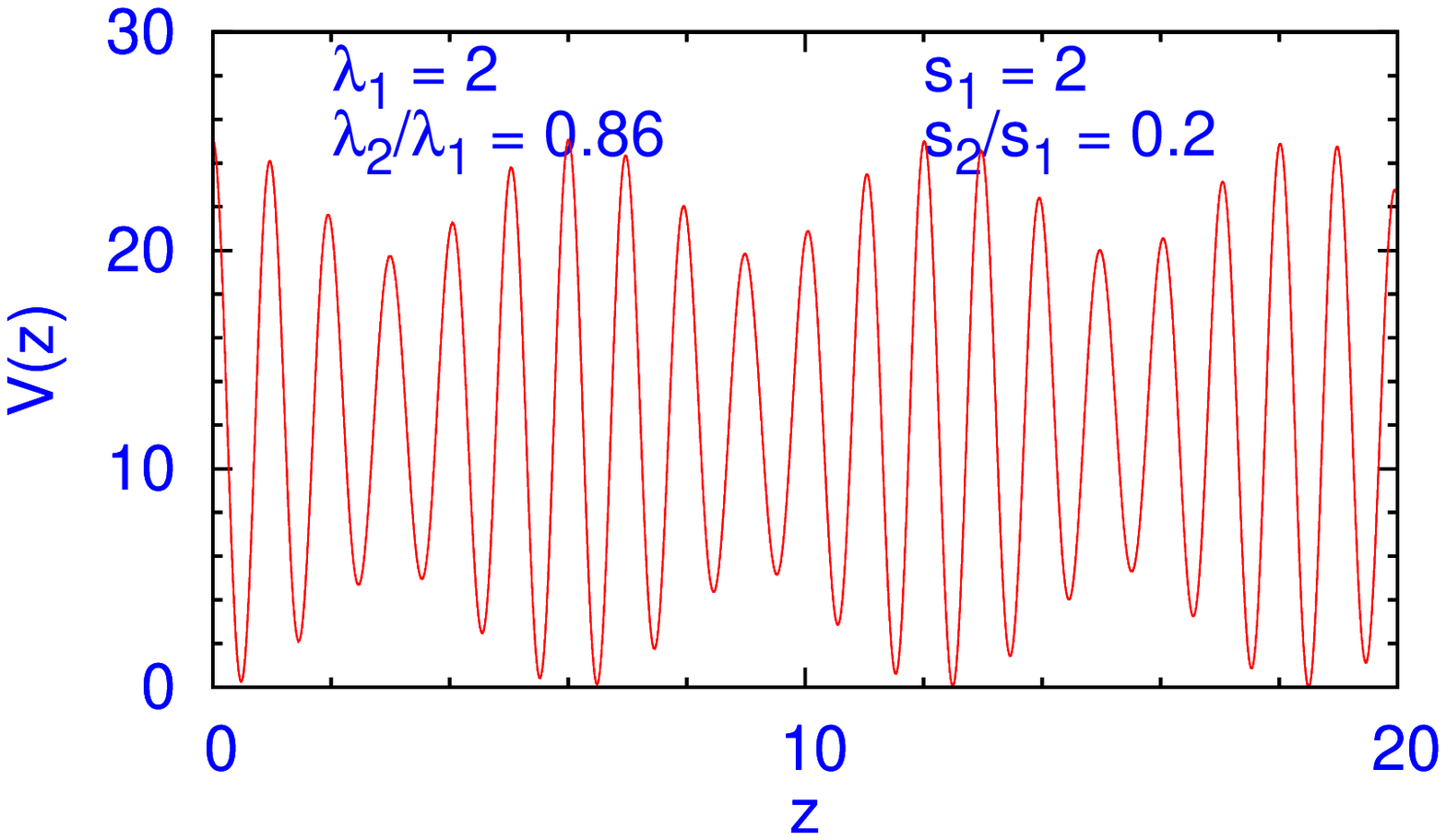}
(a)
\includegraphics[width=.9\linewidth]{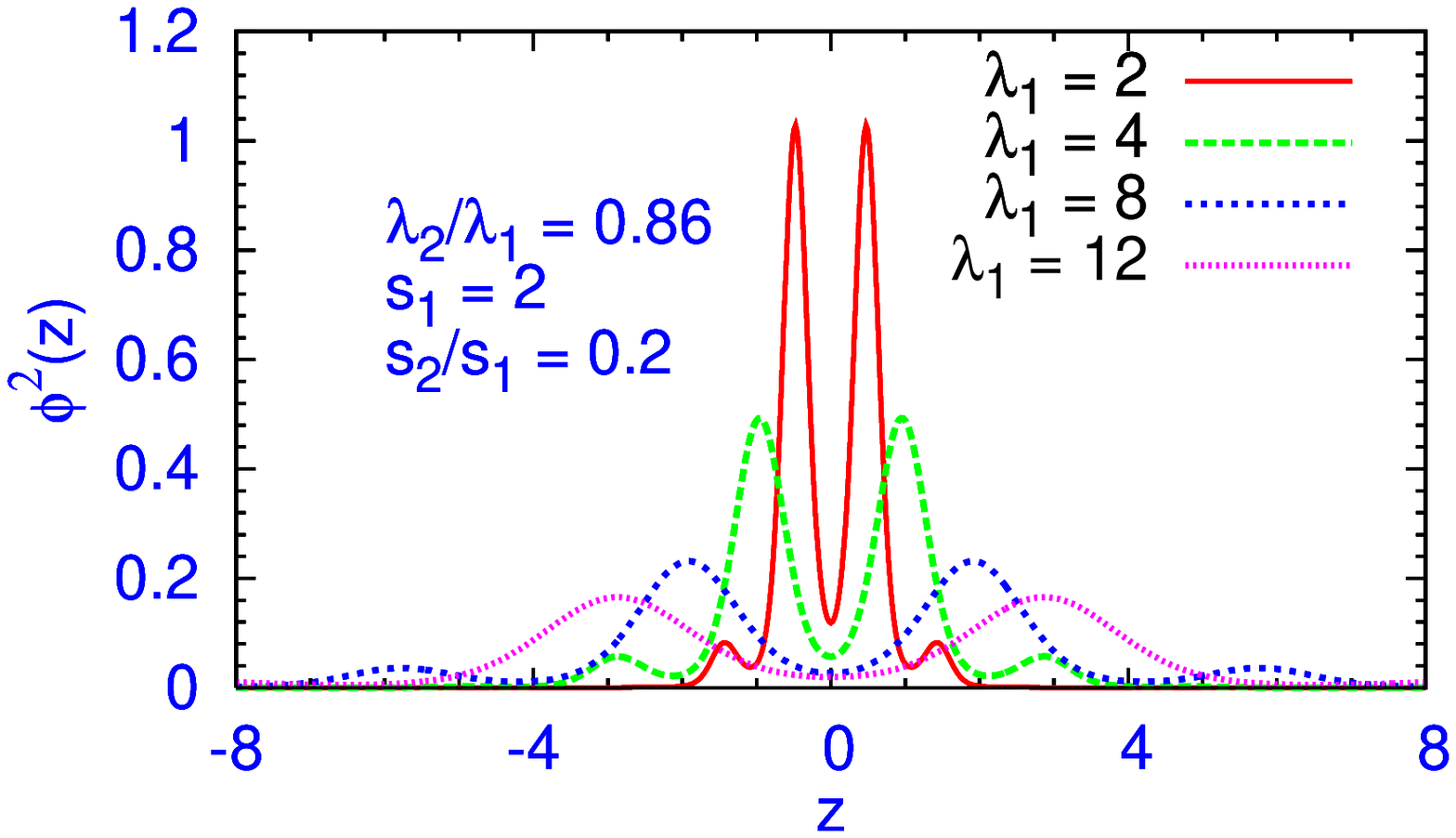}
(b)
\includegraphics[width=.9\linewidth]{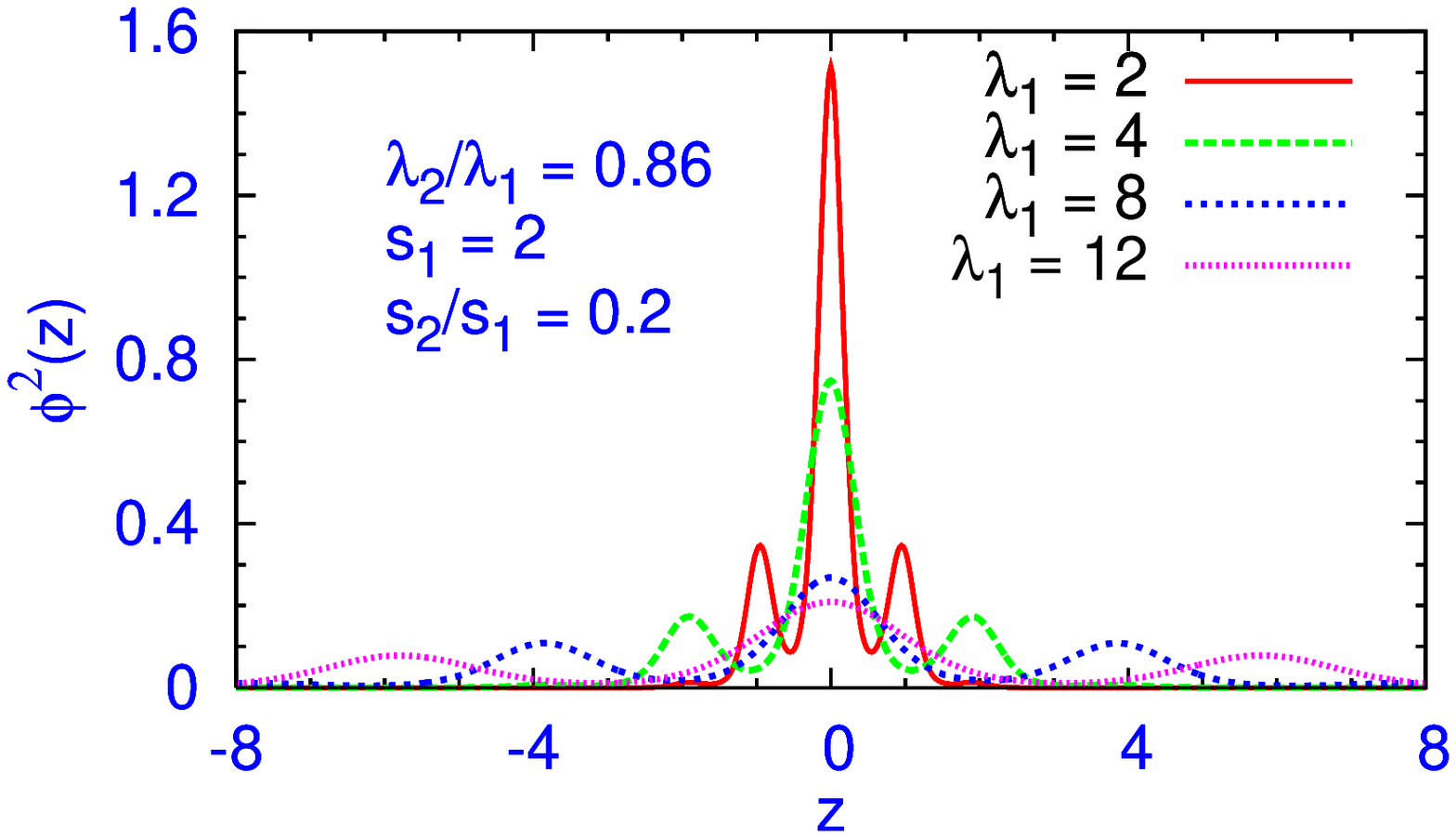}
(c)
\end{center}
\caption{(Color online)  (a) Potential $V(z)$ vs. $z$ given by Eq. 
(\ref{pot1}) for $\lambda_1=2, \lambda_2/\lambda_1=0.86, s_1 =2, $ and $s_2/s_1=0.2$ and (b) 
density distribution $\phi^2(z)$ vs. 
$z$ of a non-interacting BEC calculated with this potential
for $s_1=2$, $s_2/s_1=0.2, \lambda_ 2/\lambda_1=0.86$ and 
$\lambda_1=2,4,8, $ and 12.   (c)  
  Density distribution $\phi^2(z)$ vs. $z$ of a 
non-interacting BEC for potential (\ref{pot2}) for  $s_1=2$, 
$s_2/s_1=0.2,\lambda_2/  \lambda_ 1=0.86$ and $\lambda_1 =2,4,8,12.$ 
All  variables are expressed in dimensionless units.}
\label{fig2}
\end{figure}

Because of the oscillating nature of the potential great care was needed 
to obtain a precise { localized state}. 
The accuracy of the numerical 
simulation was tested by varying the space and time steps as well as the 
total number of space steps. A larger value of the ratio $s_2/s_1$ gives 
more binding for the localized state, consequently, the localized state 
has smaller spatial extension. 

To understand the nature of these localized states and their behavior 
under the variation of different parameters, first we consider the 
localized states with larger values of $\lambda_1$.  Such states with 
{\it large} $s_2/s_1 (=1)$ occupy a small number of OL sites and hence their 
simulation can be performed relatively easily. The shape of the 
localized state then becomes a quasi-Gaussian for potential (\ref{pot2}) 
and we compare our numerical results with the variational results in 
this case.  In Fig. \ref{fig1} (a) we plot the results of density 
distribution from numerical and variational calculations for potential 
(\ref{pot2}) for different $\lambda_1$ for fixed $s_1,s_2$, and 
$\lambda_2/\lambda_1$. In Fig. \ref{fig1} (a) real- and imaginary-time 
propagation routines produced identical results. In Fig. \ref{fig1} (b) 
we plot the density distribution $\phi^2(z)$ vs. $z$ for potential 
(\ref{pot1}) for the same set of parameters as in Fig. \ref{fig1} (a). 
In this case one has two peaks in the density distribution. We also 
calculated the energies of these states. The energies for 
$\lambda_1=15,10,$ and 5 for the potential (\ref{pot2}) of Fig. 
\ref{fig1} (a) are 0.264, 0.599 and 2.374 to be compared with the 
variational results of Eq. (\ref{energy}): 0.266, 0.594, and 2.396, 
respectively. This 
agreement between the numerical and variational results of density 
for potential
(\ref{pot2})
in 
Fig. \ref{fig1} (a), 
 and of the respective energies,  provides assurance about the accuracy of the numerical code 
used in simulation in our investigation. We also calculated the 
(numerical) energies of the density profiles displayed in Fig. 
\ref{fig1} (b) which are, respectively, 0.270, 0.603, and 2.412. These 
energies for potential (\ref{pot1}) are distinct from those of potential 
(\ref{pot2}); but the two sets of energies are very close to each other.

\begin{figure}
\begin{center}
\includegraphics[width=\linewidth]{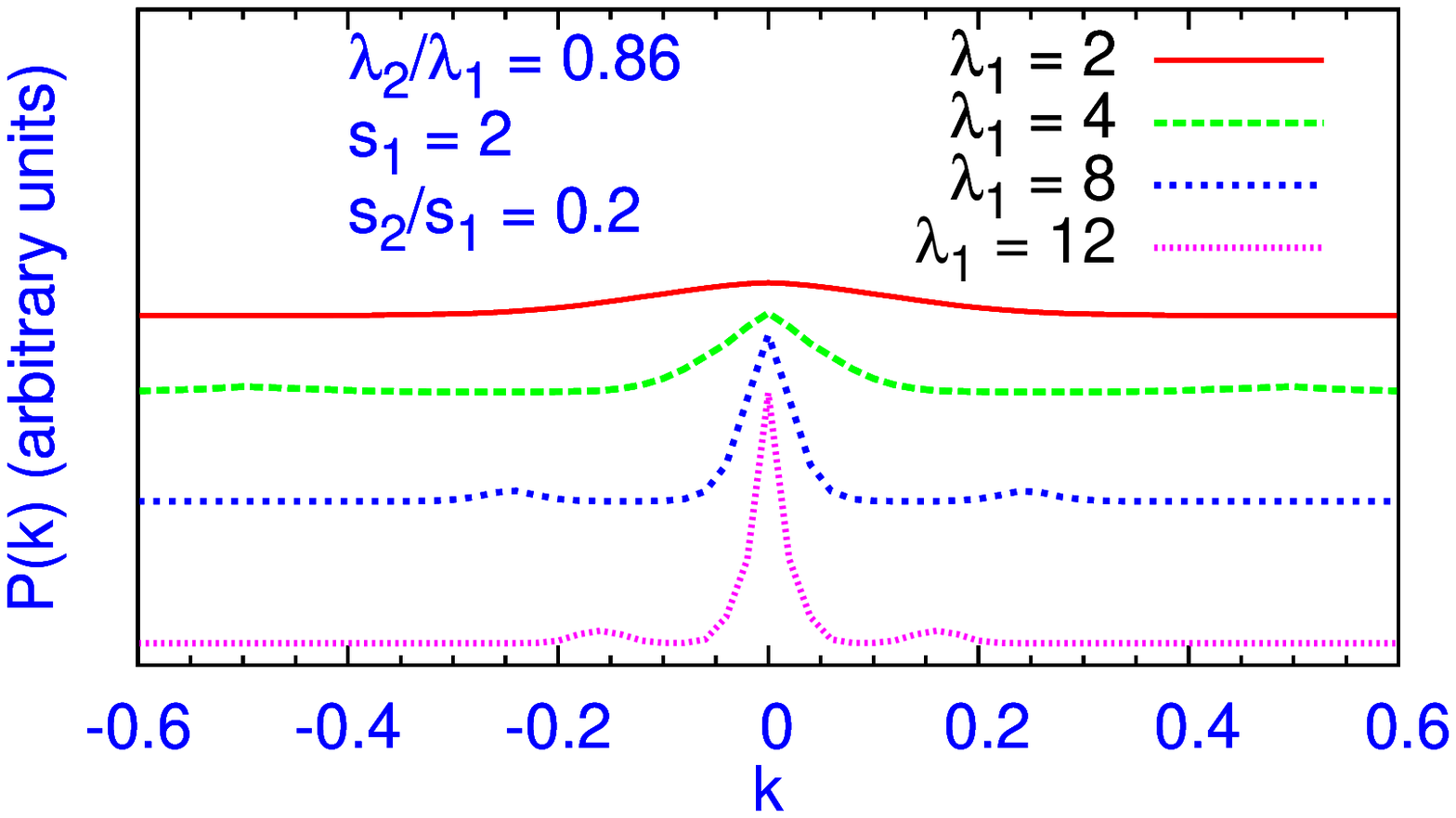}
(a)
\includegraphics[width=\linewidth]{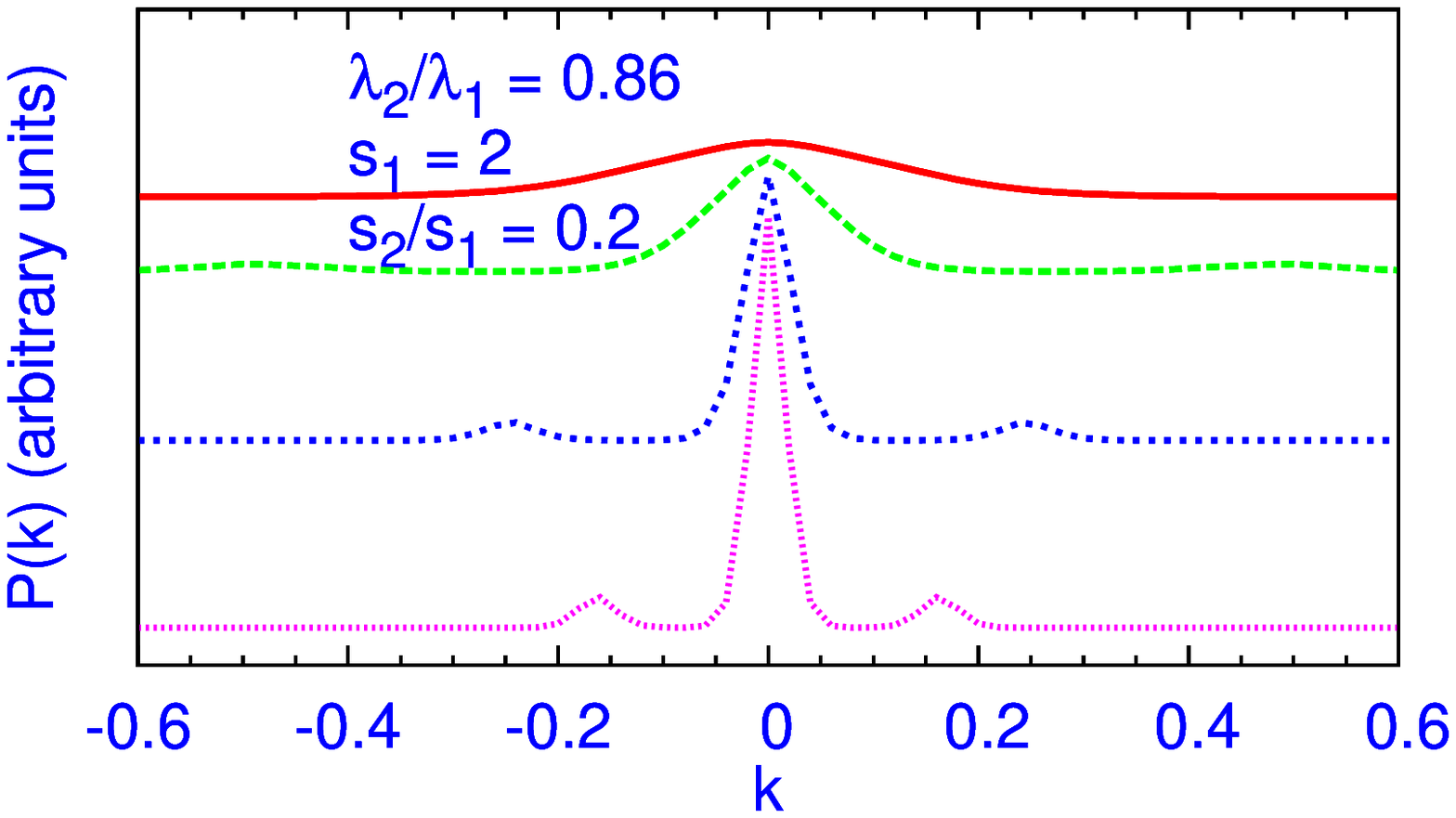}
(b)
\end{center}

\caption{(Color online) Momentum distribution $P(k)$ vs. $k$ of the
localized
states, shown in Figs. \ref{fig2} (b) and (c), for (a) potential
(\ref{pot1}) and (b)
potential (\ref{pot2}), for $\lambda_2/\lambda_1=0.86, s_1=2,s_2/s_1=0.2,$ and 
$\lambda_1=2,4,8,12$.
}

\label{fig3}
\end{figure}

\begin{figure}
\begin{center}
\includegraphics[width=\linewidth]{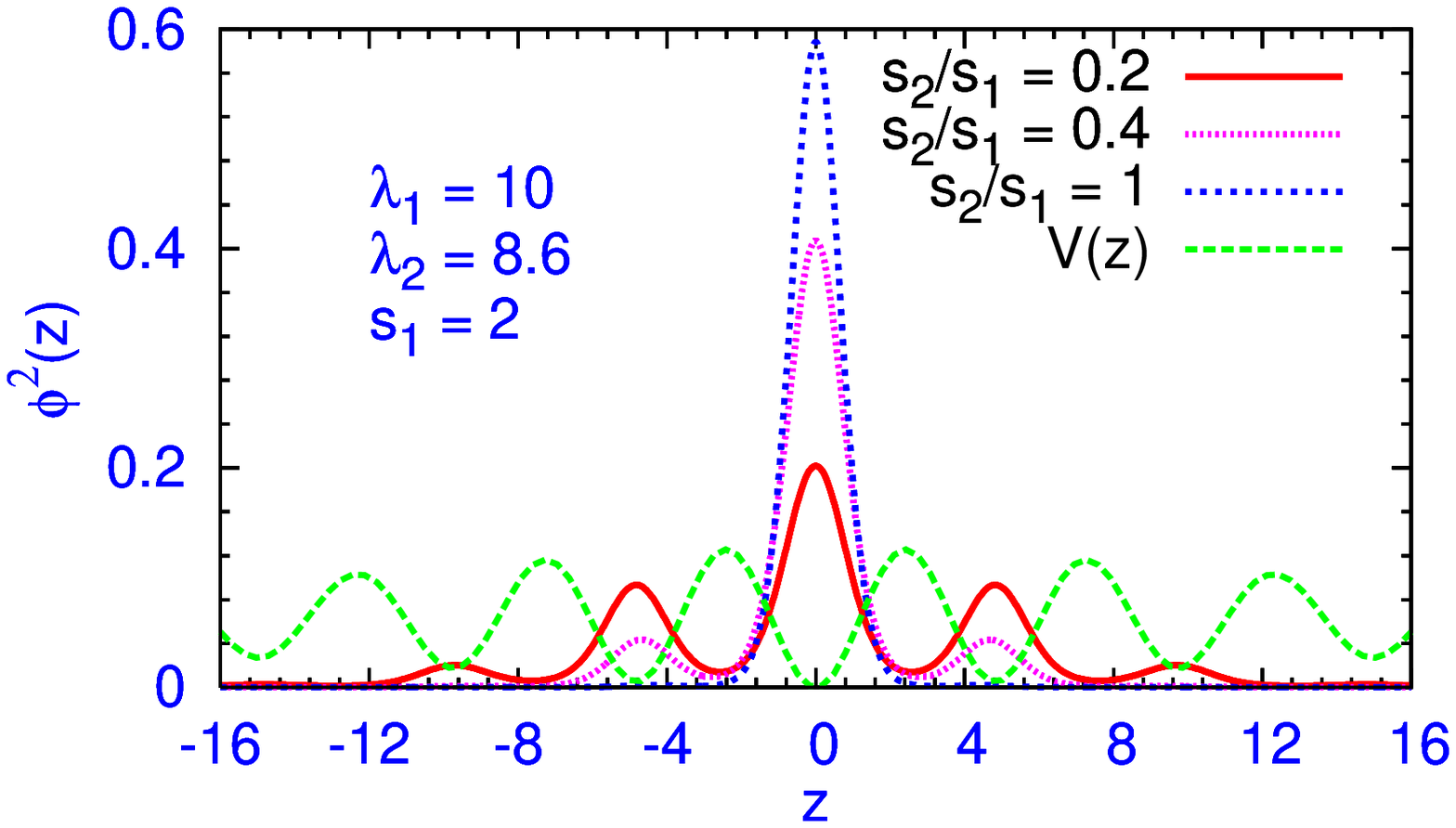}
(a)
\includegraphics[width=\linewidth]{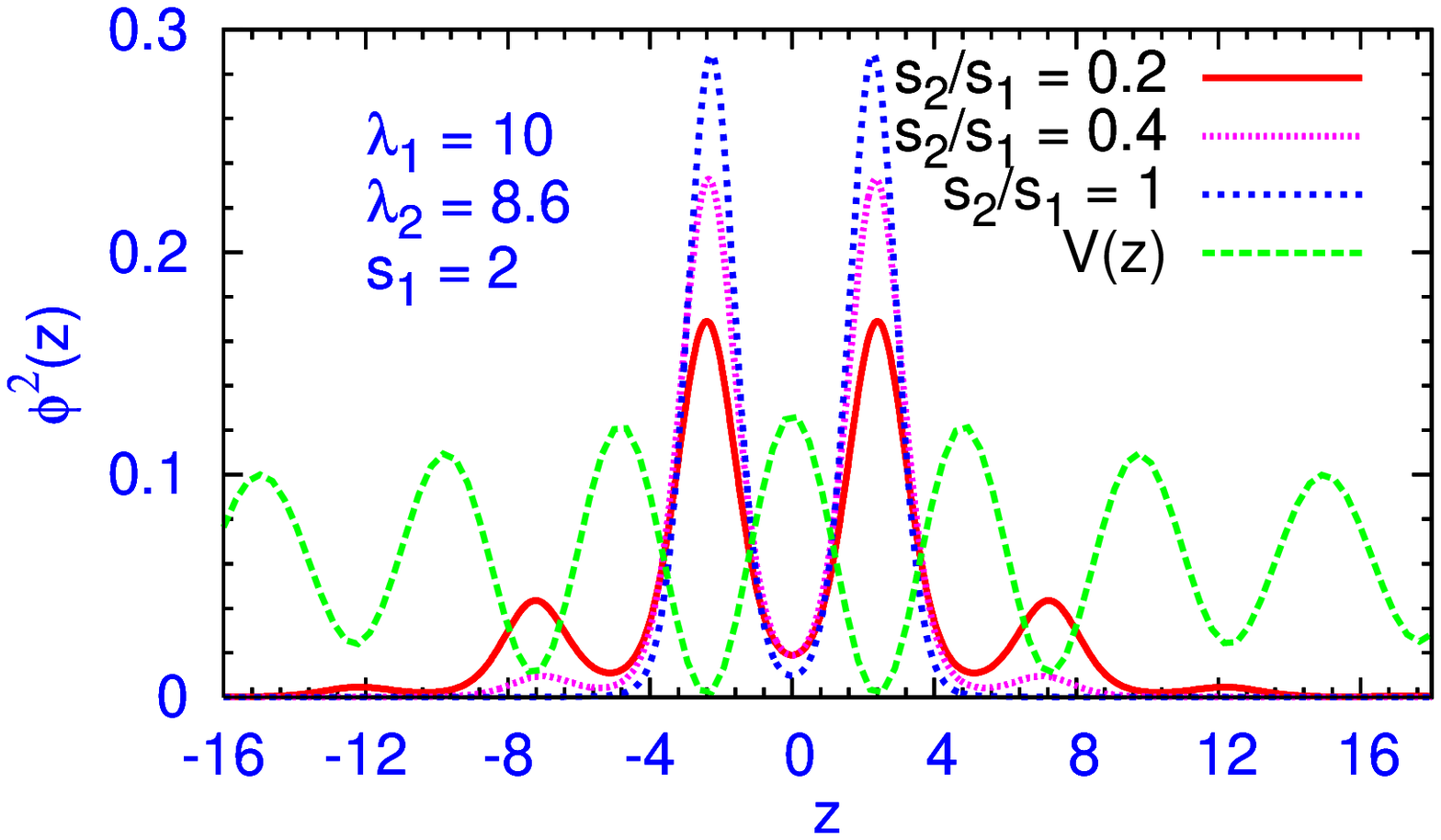}
(b)
\end{center}

\caption{(Color online) Typical density distribution $\phi^2(z)$ vs. 
$z$ for   a non-interacting BEC for 
(a) potential (\ref{pot2}) and (b)  potential (\ref{pot1})
for $\lambda_1=10, \lambda_2=8.6, s_1=2, $ 
and $s_2/s_1= 0.2, 0.4$ and 1.
The quasi-periodic OL potential $V(z)$ for $s_2/s_1=0.2$  is also 
plotted 
in 
(a) and (b) in arbitrary units.  
All variables    are   in dimensionless 
units. 
}

\label{fig4}
\end{figure}

Now  we present  the results for the solution of 
Eq. (\ref{gp}) with potential (\ref{pot1}) for  a {\it small}  $s_2/s_1$.
 The potential $V(z)$ given by 
(\ref{pot1})
 for 
$\lambda_1=2, \lambda_2/\lambda_1=0.86, s_1=2$ and $s_2/s_1=0.2$ 
is plotted in Fig. \ref{fig2} (a). 
This potential is quite similar to the potential
illustrated in Fig. 1(a) of Roati {\it et al.} \cite{roati}. 
The density 
distribution 
$\phi^2(z)$ vs. $z$ 
corresponding to the localization for this
potential is plotted in Fig.  \ref{fig2} (b) for 
$s_1=2, s_2/s_1=0.2, \lambda_2/\lambda_1=0.86$ and $\lambda_1=2,4,8, $ and 12.
It is seen that with a decrease of $\lambda_1$ more attraction is created. 
Consequently, the
states with $\lambda_1=2$ and 4 are more localized in space with lesser
spatial extension. But with further increase of $\lambda_1$ a single
site of the quasi-periodic OL  potential occupies a large
region in space. When this happens the spatial size of the 
localized bound states increases with $\lambda_1$, as the localized 
state cannot occupy less than two sites of the 
quasi-periodic OL 
potential (the density for potential (\ref{pot1})
has to be symmetric around $z=0$ and must have 
a minimum at $z=0$).  
This happens for
$\lambda_1=8 $ and 12. For $\lambda_1=8 $ and 12 the localized
bound states occupy 4 sites of the quasi-periodic OL 
potential. 
Partially  similar feature is also exhibited by the 
localized 
states of the quasi-periodic OL potential (\ref{pot2}) for 
larger values 
of $\lambda_1$ as shown in Fig. \ref{fig2} (c), where we plot 
the density distribution $\phi^2(z)$ vs. $z$ for 
$s_2=2, s_2/s_1=0.2, 
\lambda_2/\lambda_1=0.86$ and $\lambda_1 = 2,4,8, $ and 12.
As expected,  the localized state with 
potential  (\ref{pot1}) has a minimum at $z=0$, whereas the localized 
state of potential  (\ref{pot2}) has a maximum at $z=0$.  
 We plot in Figs. 
\ref{fig3} (a) and (b) the momentum distribution $P(k)
$ vs. $k$ 
for the 
localized states $-$  a quantity of experimental interest \cite{roati} $-$ 
shown in Figs. \ref{fig3} (a) and (b) for 
potentials (\ref{pot1}) and (\ref{pot2}), respectively. 
In general, as expected, the central peak of momentum distribution 
of the localized state is 
sharper for a localized state of larger spatial extension.


Now we study how these localized states are affected by a variation of 
the ratio  $s_2/s_1$ when $\lambda_1,\lambda_2$ and $s_1$ are maintained 
constant. 
For  this purpose we plot in Figs. \ref{fig4} (a)
and (b) 
 the density distribution of the localized states for potentials 
(\ref{pot2}) and (\ref{pot1}), respectively, for $\lambda_1=10, 
\lambda_2 = 8.6, s_1 =2$ and for different values of the fraction 
$s_2/s_1 = 0.2, 0.4$ and 1. In addition we plot the quasi-periodic OL 
potential $V(z)$ for $s_2/s_1=0.2$ in arbitrary units, just to compare 
the position of the maxima and minima of the potential with the position 
of the minima and maxima of density. (The position of the maxima and 
minima of the potential $V(z)$ does not change visibly with $s_2/s_1$.) 
It is found in both cases that the states with larger values of the 
fraction $s_2/s_1$ have smaller spatial extension corresponding to 
larger attraction.  The results reported in Figs. \ref{fig4} (a) and (b) 
are in qualitative agreement with a conclusion of the study of Modugno 
\cite{modugno}, that the localization appears and increases as the 
disorder to hopping ratio
$\Delta/J$  increases. The increase of 
$\Delta/J$ is related to an increase of $s_2/s_1$ for a fixed 
$\lambda_1$ and $\lambda_2$, exactly as illustrated in Figs.  \ref{fig4} 
(a) and (b). However, because of the distinct model (DNLSE) used in Ref. 
\cite{modugno}, in contrast to the numerical solution of the GP equation 
in the present study, a quantitative comparison of the results of the two 
studies is not to the point.

\begin{figure}
\begin{center}
\includegraphics[width=\linewidth]{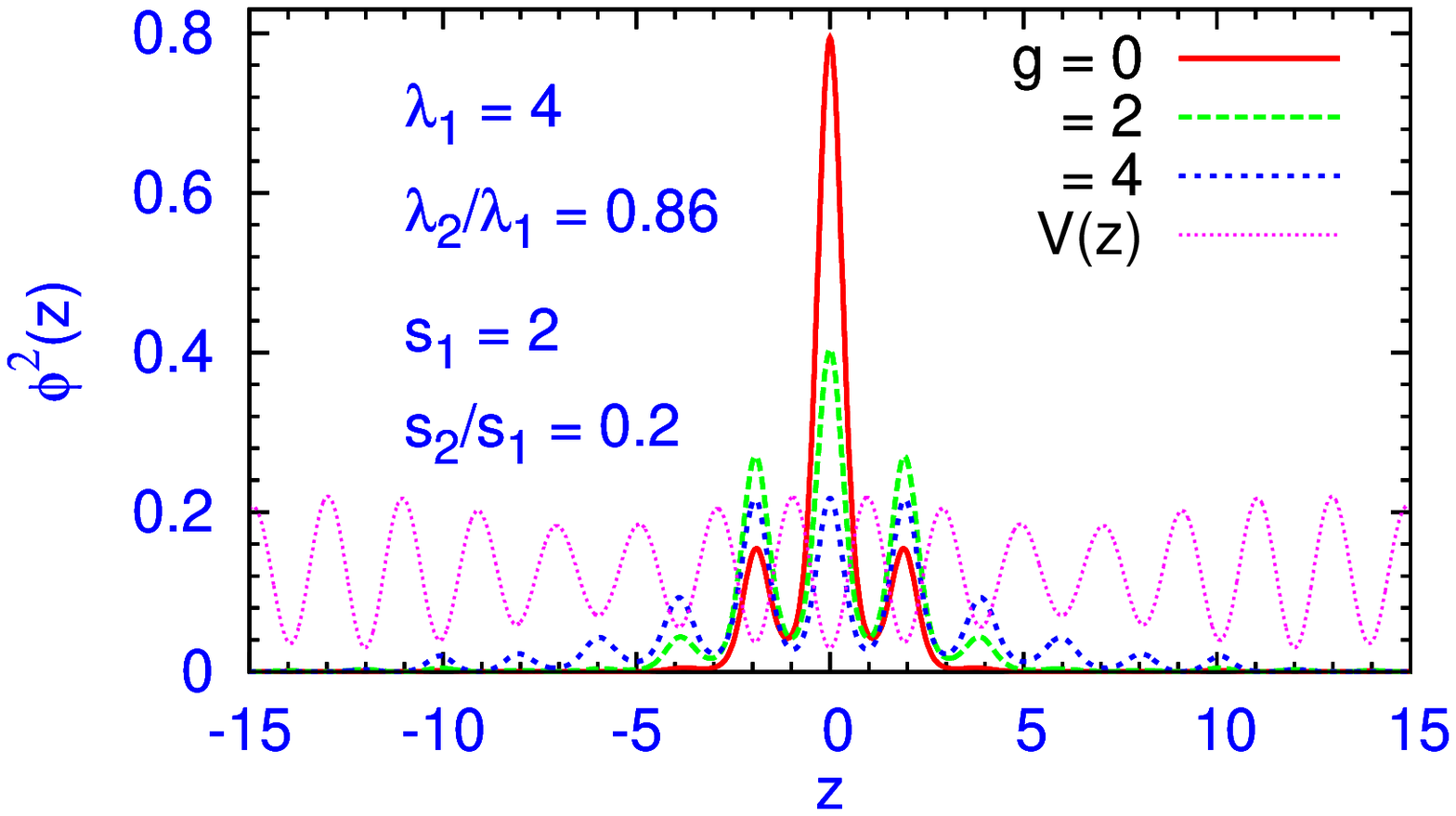}
(a)
\includegraphics[width=\linewidth]{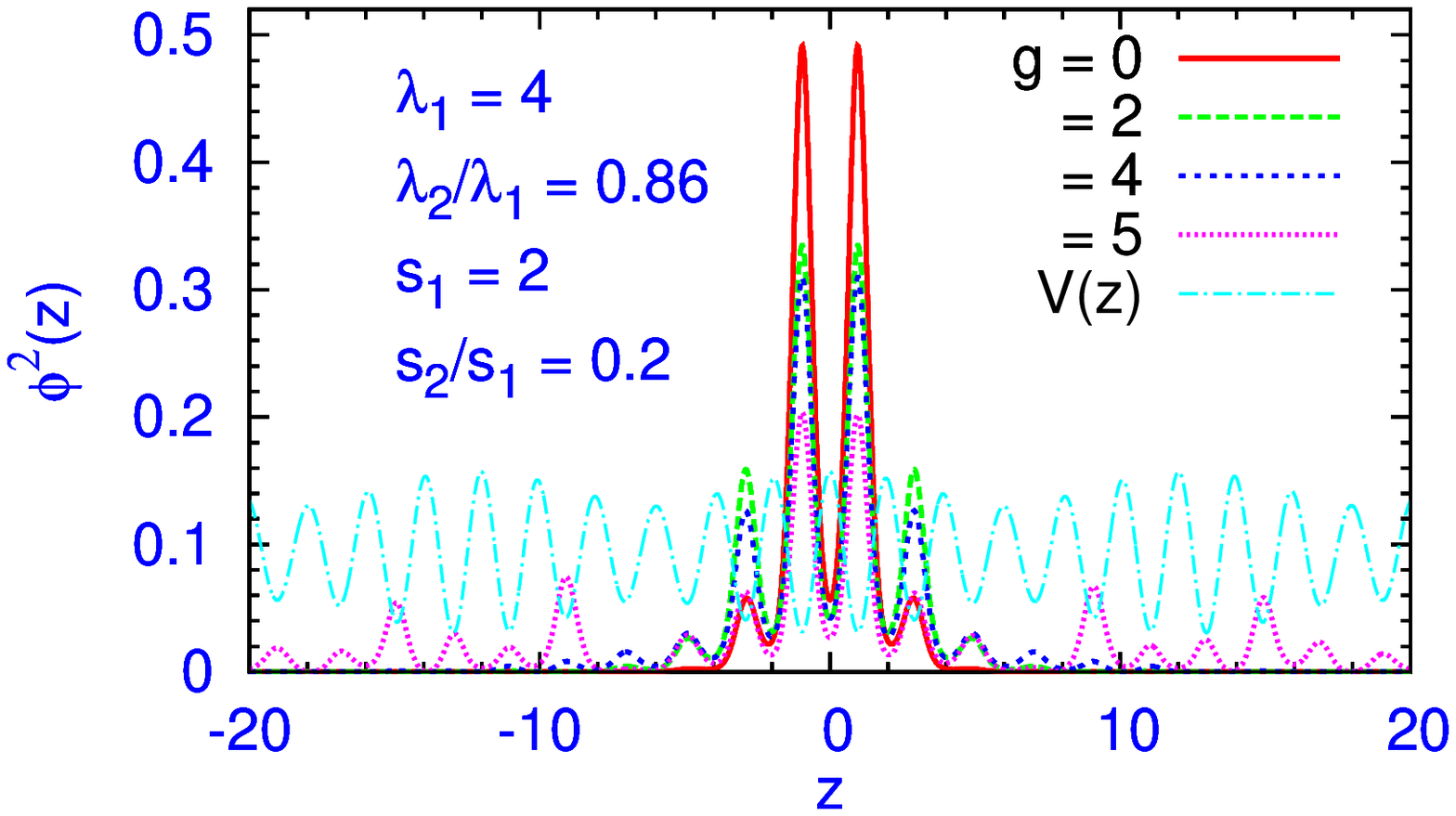}
(b)
\includegraphics[width=\linewidth]{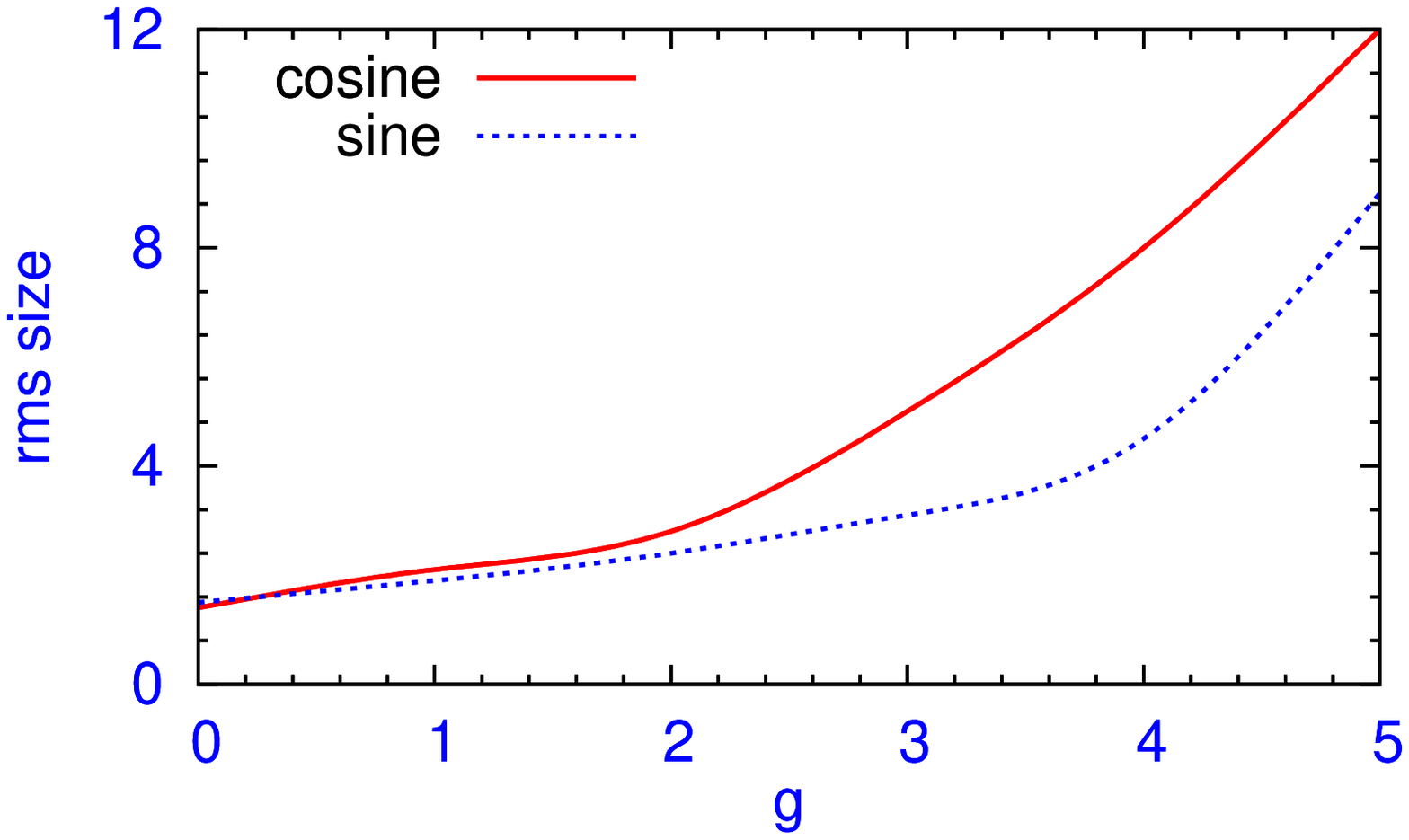}
(c)
\end{center}

\caption{(Color online) Typical density distribution $\phi^2(z)$ vs. 
$z$ for   an interacting  BEC for  different $g\equiv 2N\hat a/a_\perp$ for  
(a) potential (\ref{pot2}) 
and (b) potential (\ref{pot1}) 
for $\lambda_1=4, \lambda_2/\lambda_1=0.86, s_1=2, s_2/s_1=0.2 
$.  The quasi-periodic OL potential $V(z)$ is plotted in arbitrary units.
(c) The rms size vs. non-linearity $g$ of the stable condensate in the 
quasi-periodic potential (\ref{pot1}) (cosine) and  (\ref{pot2}) (sine)
for $\lambda_1=4, \lambda_2/\lambda_1=0.86, s_1=2, s_2/s_1=0.2.$
All quantities plotted are in  dimensionless units. 
}
\label{fig5}
\end{figure}

\begin{figure}
\begin{center}
\includegraphics[width=\linewidth]{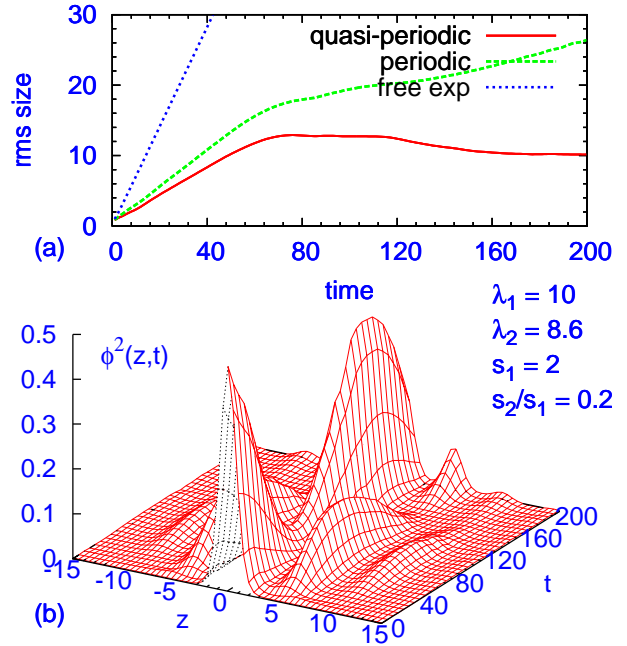}
\end{center}

\caption{(Color online) (a) The rms size  of an 
expanding  BEC 
released from a harmonic trap into the quasi-periodic OL  potential 
(\ref{pot2}) with $\lambda_1=10, \lambda_2 =8.6, s_1 =2, s_2/s_1=0.2$  
at time $t=0$. The rms sizes during expansions in a periodic OL 
potential with $s_2=0$  
and free expansion are 
also shown.   (b) The density profiles during above expansion in the 
quasi-periodic OL  potential
are illustrated by plotting $\phi^2 (z,t)$ vs. $z$ and $t$.  
}
\label{fig6}
\end{figure}

Next we study the effect of including interaction in a BEC of $^{39}$K 
atoms with scattering length $\hat a= 33a_0=1.75$ nm \cite{wang} (with 
$a_0=05292$ nm, the Bohr radius,) by solving Eq. (\ref{gp}) with 
potentials (\ref{pot1}) and (\ref{pot2}). In present dimensionless units 
this will correspond to a scattering length of 
$a\equiv \hat a/a_\perp= 0.00175$. The 
inclusion of the repulsive non-linear potential term in Eq. (\ref{gp}) 
will reduce the possibility of the appearance of localized bound states. 
This is illustrated in Figs. \ref{fig5} (a) and (b) where we plot the 
density distribution for $\lambda_1=4, \lambda_2/\lambda_1=0.86, s_1=2, 
s_2/s_1=0.2$ for potentials (\ref{pot2}) and (\ref{pot1}), respectively, 
for different $g\equiv 2N\hat a/a_\perp=(0,2,4,5)$. The corresponding 
quasi-periodic OL potentials are also plotted in arbitrary units. (The 
advantage of using the variable $g$, rather than the scattering length 
$\hat a$ and number of atoms $N$, in these plots is that the present plots 
can easily be used to simulate different experimental situations with 
different traps and distinct bosonic atoms.) For $g=0$, for both 
potentials the localized states are confined between $z=\pm 10$.  For 
$g=2$, with increased repulsion, the matter density is reduced in the 
central peaks and new peaks appear for larger $z$ values. For $g=4$, 
with further increase in repulsion the matter density is further reduced 
in the central region and new peaks appear in the form of ondulating tails
near the edges.  With further 
increase in the value of $g$, the localized states have larger and 
larger spatial extension and soon the non-linear repulsion is so large 
that no localized states are possible and this happens rapidly as $g$ is 
increased beyond 6. The non-linearity in Eq. (\ref{gp}) is $g=2 \hat a 
N/a_\perp$ and for about 1800 $^{39}$K atoms with $a=0.00175$ 
\cite{wang} the non-linearity has the typical numerical value $g\approx 
6$. Such a small non-linearity can have a large effect on localization 
of a $^{39}$K BEC and can prohibit the localization. However, the number 
of K atoms can be proportionately increased if the scattering length is 
reduced by varying an external background magnetic field near a Feshbach 
resonance \cite{fesh}. 
As $g$ value is increased, the root mean square 
(rms) size of the BEC increases before reaching a critical $g$ value 
($>6$) corresponding to the destruction of localization. 
(It is difficult to obtain accurately the critical value
 of $g$ needed to destroy the localization.)
The increase in 
the rms size of the localized state with the increase in $g$ is 
illustrated in Fig. \ref{fig5} (c) where we plot the rms size vs. $g$ 
for potentials (\ref{pot1}) and (\ref{pot2}).
It should be noted that in the experiment of 
Roati {\it et al.} \cite{roati} the residual scattering length of 
$^{39}$K atoms near the Feshbach resonance 
was 0.1$a_0$ (= 0.0053 nm), 
e.g., they can vary the scattering length in such small steps. 
Thus it should be possible experimentally to obtain the 
curves illustrated in Fig.  \ref{fig5} (c) and compare them  with the 
present investigation.

So far we studied the stationary properties of the  localized 
state. Next we study some dynamical aspects of the localization. 
One interesting feature is what happens when a BEC is released from a 
harmonic trap into a quasi-periodic OL trap as investigated in the 
experiment of 
Billy {\it et al.} \cite{billy}. Mathematically, it means 
that at time $t=0$  the harmonic trap is suddenly changed into a 
quasi-periodic OL trap. The evolution of the rms size  of the BEC with 
time is plotted in Fig. \ref{fig6} (a) for potential (\ref{pot2}). The 
rms size first increases 
with the expansion of the BEC and after a certain amount of expansion it 
will be locked in an  appropriate  localized state. After this 
happens the system executes breathing oscillation around a mean shape of 
the localized state and the rms size remains bounded and does not 
increase indefinitely with time. To compare with this behavior we also 
plotted the rms size for expansion in a pure periodic OL potential when 
the disorder is removed by setting $s_2$ =0 in  potential (\ref{pot2}). In 
that case there is no localized state and the system expands 
for ever with an ever-increasing rms size. Finally, when all OL 
potentials are removed by setting $s_1=s_2=0$ the system increases 
monotonically with a higher expansion rate.  The dynamical density 
profile of the 
BEC  during the expansion and locking in a  localized state for 
the  quasi-periodic potential (\ref{pot2}) is shown in Fig. \ref{fig6} 
(b) where the initial expansion until $t=80$ and the consequent 
breathing oscillation of the BEC is clearly illustrated. 

\section{Conclusion}

In this paper, using the numerical solution of the GP equation, we 
studied the localization of a non-interacting BEC in a quasi-periodic 1D 
OL potential prepared by two overlapping polarized standing-wave laser 
beams with different wavelengths and amplitudes. Specifically, we 
considered two analytical forms (sine and cosine) of the OL potential. 
We studied the effect of the variation of wavelengths and amplitudes on 
the localization. We also studied the non-linear dynamics when a BEC is 
released from a 1D harmonic trap into a quasi-periodic 1D OL trap. After 
release, the BEC first expands (from the tightly bound harmonic 
oscillator bound state) and then the expansion is stopped and the BEC is 
found to be trapped into one of the localized states in the 
quasi-periodic OL potential, as observed by Billy {\it et al.} \cite{billy}.

We also studied the effect of a repulsive atomic interaction on 
localization. It is found that a repulsive atomic interaction destroys 
the localization. We studied the route to this destruction of 
localization in some details. In particular, we investigated the 
localization as the non-linearity $g\equiv 2N\hat a/a_\perp$ of the 
non-linear 1D GP equation is increased. 
It is found that as  $g$ is gradually increased, the localization is slowly 
weakened with the localized state extending over a large space domain. 
Eventually, for $g$ greater than about 6 or so, the localization is destroyed.

There have been previous studies of some aspects of the localization of 
a BEC in a quasi-periodic potential \cite{modugno} and also its 
destruction \cite{dnlse}. (It should be noted that the present study is 
mostly complimentary to these previous studies, rather than 
overlapping.) However, in the present study we consider a direct 
numerical solution of the GP equation as opposed to a solution of the 
DNLSE used in the previous studies. In view of the rapidly oscillating 
nature of the quasi-periodic OL potential and 
of the solution of the non-linear equation, the 
results of the direct numerical solution of the GP equation, as used in 
the present investigation, { should be}  more useful for a direct 
comparison with the experiments. 

\label{IIII}

\acknowledgments

FAPESP (Brazil),  CNPq (Brazil) and CARIPARO (Italy)
provided partial support.

\end{document}